# Theoretical Modeling of Structure-Toxicity Relationship of Cyanides

**Marcin Molski**

*Quantum Chemistry Department, Faculty of Chemistry,*

*Adam Mickiewicz University of Poznań*

*ul. Uniwersytetu Poznańskiego 8, 61-614 Poznań, Poland*

e-mail: mamolski@amu.edu.pl

## Abstract

The global descriptors of chemical activity: ionization potential IP, electron affinity EA, chemical potential μ, absolute electronegativity χ, molecular hardness η and softness S, electrophilicity index ω for cyanides $X(CN)_k$ with X=H, Na, K, Ag, Cu, Ca, Hg, Cd, Zn in the gas phase and water medium have been determined by taking advantage of the quantum-chemical computations. To this aim, the HOMO and LUMO energy levels were calculated using DFT B3LYP method and QZVP (Valence Quadruple-Zeta Polarization) basis set, which enables precise calculations for hydrogen cyanide and its salts containing both light (H, Na, Ca) and heavy (K, Ag, Cu, Cd, Hg, Zn) atoms. The results obtained indicate that while the EA-parameter roughly determines the $LD_{50}$ values for the cyanides considered, the ω-descriptor is related rather to the product of cyanide $LD_{50}$ and hydrolysis n-degree. Hence, the theoretical $LD_{50}^{CN}(\omega)$ function proposed is interconnected with the $n \cdot LD_{50}$ collective variable, whereas $LD_{50}(EA)$ directly approximates $LD_{50}$ values indicating that the toxicity of cyanides decreases with increasing EA, ω-values. The calculations carried out suggest that some of the $LD_{50}(Exp)$ experimentally determined are incorrect (AgCN) or inaccurate ($Cd(CN)_2$) and require revision. Comparison of the theoretically estimated $LD_{50}(\omega)$ with NOAEL toxicity parameters indicates that they are well correlated in contrast to $LD_{50}(EA)$ and $LD_{50}(Exp)$, exhibiting lower and marginal degrees of correlation, respectively.

## 1. Introduction

Theoretical toxicity prediction is a challenging problem that needs rapid and efficient methods for *in silico* modeling of bioactive chemical components of drugs, supplements and cosmetics (Arwa and Vladimir, 2016; Dearden, 2003; Deeb and Goodarzi, 2012; Kleandrova et al., 2015; Valerio, 2009). This is an important issue, both from an economic point of view - *in silico* studies are cheaper than experimental ones and reduce the high cost of clinical trials as well as the ethical one, associated with the possibility of reducing, and ultimately eliminating, tests conducted on animals. For this purpose, descriptors of chemical reactivity are sought, which enable the quantitative description of the toxicity of substances, both known and modeled by *in silico* techniques. Investigations on quantitative structure-activity relationships (QSARs) carried out in recent years (Cherkasov et al., 2014; Croni, 2002; Guha and Willighagen, 2012; Karelson, 2000; Todeschini and Consonni, 2009) provided numerous activity parameters, quantifying various electronic, geometric, topological or steric properties of molecules, which can be classified as global or local ones. In the first case, descriptors are computed for the system as a whole, whereas in the second one, only for its individual fragments, having the biggest impact on the observed bioactivity. The simplest global descriptors are: ionization potential IP, electron affinity EA, chemical potential $\mu$, absolute electronegativity $\chi$, molecular hardness $\eta$ and softness S, electrophilicity index $\omega$, which have been successfully employed in describing the chemical activity in general and toxicity in particular (Al-Fahemi 2012; Marinho et al., 2021; Talmaciu et al. 2016). To evaluate the parameters mentioned, we need only the values of energies of the highest occupied (HOMO) and lowest unoccupied (LUMO) molecular orbitals to be calculated using quantum-chemical computational methods implemented in commonly available software packages. In view of this, the main purpose of the present study is to check which of the parameters mentioned can be applied in the quantitative description of the toxicity of hydrogen cyanide and its salts. The simplicity of their structures enables high-accuracy quantum-chemical calculations with the use of extensive functional basis sets and advanced computational methods. The studies conducted so far on organic compounds have revealed that their toxicity is correlated with the electrophilicity index $\omega$ (Parthasarathi et al., 2004; Roy, 2006; Shalini, 2017) and the energy of the HOMO orbitals (Vikas, 2015). Therefore, confirmation of the usefulness of the global parameters mentioned for description of the toxicity of inorganic substances will make them a universal, powerful tool in the context of QSAR investigations. An additional goal of the present study is the verification of the experimental $LD_{50}$ values reported for hydrogen cyanide and its salts as part of them is inconsistent and significantly differ from each other, even if they were determined for the same animal species using an identical administration method. We shall also be concerned with the comparison of the theoretically generated $LD_{50}$ values with the more advanced toxicity parameters NOAEL (No Observed Adverse Effect Level) and AEGL (Acute Exposure Guideline Level).

## 2. Materials and methods

Among the numerous experimental methods for quantifying the acute toxicity of substances is the determination of the median lethal dose $LD_{50}$ for a toxin, which is required to kill 50% of a group of experimental animals after a specified duration of exposure. This method provides results depending on animal species and mode of administration: inhalation, intraperitoneal, intravenous, intramuscular, oral, subcutaneous. However, even in the case of the same animal and administration procedure, the $LD_{50}$ may vary due to different factors such as: the genetic characteristics of the population tested, age, sex, diet (restricted or standard) or when the determinations are carried out in different laboratories (Egekeze and Oehme, 1980; Lorke, 1983). In Table 1 the experimental values of $LD_{50}$ for HCN and its salts, obtained by peroral exposure to rats are presented. Its analysis reveals that the toxicological characteristics of cyanides may be ambiguous and seriously differ in dependence on the source. In such circumstances, we have decided to use for comparison of the experimental and theoretical $LD_{50}$ for HCN, NaCN and KCN those evaluated by Ballantyne and Salem (2008) for female, unstarved rats and oral administration. They seem to be the most reliable as they were determined with the high precision at 95% Confidence Limit CL: $LD_{50}$=4.21 for HCN CL=<3.76, 4.95>, $LD_{50}$=5.72 for NaCN CI=<5.23, 7.08> and $LD_{50}$=7.49 for KCN CL=<6.68, 8.48> in units [mg/kg BW (Body Weight)]. An additional source of $LD_{50}$ for cyanides was U.S. National Library of Medicine (NLM https://www.nlm.nih.gov). Of 19 toxic cyanide salts mentioned by Christopher and Holstege (2010), the NLM source provides $LD_{50}$ values (oral exposure to rats) for: AgCN, CuCN, KCN, NaCN, $Ca(CN)_2$, $Cd(CN)_2$, $Hg(CN)_2$, $Zn(CN)_2$. In the case of $Pb(CN)_2$, only the estimation $LD_{50}$ >1000 [mg/kg BW] is presented, whereas for HCN, NLM reports $LD_{50}$=0.81 [mg/kg BW] only for intravenous treatment. In such circumstances, those data have not been taken into consideration. Nowadays, the toxicity descriptor $LD_{50}$ has only historical meaning and is replaced by other, more reliable parameters (Howard et al., 1955; Philbrick et al., 1979, Faust, 1994, Hardy, et al., 2017), among which worth mentioning are:

- AEGL (Acute Exposure Guideline Levels) – represent the threshold exposure limit, above which the organisms tested could experience discomfort, irritation, and other non-sensory effects during exposure ranging from 10 minutes to 8 hours.
- BMD (Benchmark Dose) – amount of the toxin that produces a predetermined change in the response rate of an adverse effect, e.g., in growth, lifespan, weight, number of red blood cells of an organism tested.
- LOAEL (Lowest Observed Adverse Effect Level) – the lowest concentration of a toxin that causes adverse effects in an organism tested in comparison to normal organisms under identical conditions of exposure.

- MTD (Maximum Tolerance Dose) – which does not cause signs of poisoning, especially during the long-term investigations.
- NOEL (No Observed Effect Level) – it is the threshold dose at which any toxic symptoms are observed.
- NTEL (No Toxic Effect Level) – it represents the largest dose administered to the most sensitive species for a given duration, which produced no observed toxic effect.
- NOAEL (No Observed Adverse Effect Level) – it is the largest dose causing no tissue toxicity as well as various physiological effects of poisoning.

The values of NOAEL for hydrogen cyanide and its selected salts are reported, for example, in IRIS (Integrated Risk Information System available at https://www.epa.gov/iris), whereas the source of AEGL-1,2,3 for NaCN, KCN and Ca(CN)$_2$ is NRC (2015).

Table 1. Experimental data on acute lethal toxicity of HCN and its salts, obtained by peroral exposure to rats. Abbreviations: F – female, M – male, S – starved, US – unstarved animals.

| Compound | CAS | LD$_{50}$ [mg/kg BW] | LD$_{50}^{CN}$ [mg CN/kg BW] | Sex | Diet | Source |
|---|---|---|---|---|---|---|
| HCN | 74-90-8 | 3.62 | 3.49 | F | S | Ballantyne&Salem (2008) |
|  |  | 4.21 | 4.05 | F | US | Ballantyne&Salem (2008) |
|  |  | 4.50 | 4.33 | - |  | Forst (1928) |
| Mean |  | 4.11(37) | 3.96(70) |  |  |  |
| KCN | 151-50-8 | 5.00 | 2.00 | M |  | NLM[a], Lorke (1969) |
|  |  | 6.00 | 2.40 | M |  | Lorke (1969) |
|  |  | 7.49 | 2.99 | F | US | Ballantyne&Salem (2008) |
|  |  | 9.69 | 3.87 | F | S | Ballantyne&Salem (2008) |
|  |  | 10.00 | 4.00 | M |  | Hayes (1967) |
| Mean |  | 7.64(1.96) | 3.05(88) |  |  |  |
| NaCN | 143-33-9 | 5.09 | 2.70 | F | S | Ballantyne&Salem (2008) |
|  |  | 5.72 | 3.04 | F | US | Ballantyne&Salem (2008) |
|  |  | 6.44 | 3.42 | F |  | NLM, Ballantyne (1988) |
|  |  | 8.00 | 4.25 |  |  | Sterner (1979) |
|  |  | 15.00 | 7.96 | M |  | Smyth (1969) |
| Mean |  | 8.05(3.61) | 4.27(1.83) |  |  |  |
| Cd(CN)$_2$ | 542-83-6 | 16.00 | 5.06 |  |  | NLM |
| Hg(CN)$_2$ | 592-04-1 | 26.00 | 5.36 |  |  | NLM |
| Ca(CN)$_2$ | 592-01-8 | 39.00 | 22.03 | M |  | NLM, Smyth (1969) |
| Zn(CN)$_2$ | 557-21-1 | 54.00 | 23.93 |  |  | NLM |
| AgCN | 506-64-9 | 123.00 | 23.90 |  |  | NLM |
| CuCN | 544-92-3 | 1265.00 | 367.47 |  |  | NLM |

[a]U. S. National Library of Medicine https://www.nlm.nih.gov

### 3. Theory and calculation

An important source of information about the reactivity of chemical compounds is the difference between the energies of HOMO and LUMO orbitals. The shapes of these orbitals, as well as size of the energy gap (usually expressed in electron volts [eV])

$$\Delta E = E_{LUMO} - E_{HOMO} \quad (1)$$

have an impact on the chemical reactivity of the compound. A large energy difference defines a "hard" molecule that is more stable and less active, while a small energy gap defines a "soft" molecule that is less stable and more reactive. Based on the energy of LUMO and HOMO orbitals as well as the Koopmans' theorem for closed-shell molecules, one may define a set of parameters representing the so-called global activity descriptors (Parr and Yang,1989; Pearson,1997), which model the physicochemical properties of chemical compounds. The most important parameters of this type are:

**Ionization potential**

$$IP = -E_{HOMO} \quad (2)$$

This is the minimum energy that must be provided to detach the electron away from the molecule's HOMO orbital and bring it to infinity. The smaller IP value indicates a greater tendency of the molecule to attend the chemical reaction related to electron transfer. Vikas (2015) investigated the acute toxicity of 252 diverse organic chemicals towards *Daphnia magna* employing QSAR models based on the HOMO, total and electron-correlation energies as toxicity descriptors. He suggested that the intramolecular interactions between electrons in molecules play a vital role in the origin of acute toxicity, which is in fact an unexplored phenomenon.

**Electron affinity**

$$EA = -E_{LUMO} \quad (3)$$

It expresses the ability of a molecule to attach an electron and form an anion. EA quantifies the energy released from this process (EA> 0). Negative electron affinity (observed in the case of noble gases, nitrogen, beryllium, and magnesium) means that the energy of the anion is greater than the energy of the neutral molecule, therefore attaching an electron to it requires energy.

**Chemical hardness**

$$\eta \approx \frac{E_{LUMO} - E_{HOMO}}{2} \tag{4}$$

It determines the low susceptibility of a molecule to deformation or polarization of the electron cloud under the influence of external factors, e.g., reagents. Together with the concept of chemical softness, it forms the basis of the HSAB (Hard and Soft Acids and Bases) concept of Lewis acids and bases. When a Lewis acid reacts with a Lewis base, the electrons from the HOMO base orbital are transferred to the LUMO acid orbital. The transfer efficiency depends on the relative energies of the LUMO of the acid and the HOMO of the base and affects the stability of the reaction product. HSAB theory predicts that Lewis hard bases react preferentially with hard acids, and soft bases form stronger bonds with soft acids. The exact formula for chemical hardness is as follows

$$\eta = \left( \frac{\partial^2 E}{\partial N^2} \right)_{V(r)} \tag{5}$$

Here E stands for the total electron energy, N is the number of electrons, and V (r) is the external electrostatic potential generated by atomic nuclei that attract electrons.

**Chemical softness**

$$S \approx \frac{1}{E_{LUMO} - E_{HOMO}} \tag{6}$$

The inverse of chemical hardness - characterizes molecules with high susceptibility to deformation and polarization of the electron cloud. The precise softness formula is given by the equation

$$S = \left( \frac{\partial N}{\partial \mu} \right)_{V(r)} \tag{7}$$

dependent on the chemical potential µ, defined below.

**Chemical potential**

$$\mu \approx \frac{E_{LUMO} + E_{HOMO}}{2} \tag{8}$$

It describes the thermodynamic activity of substances and is used in the derivation of the phase equilibrium constants and chemical reactions. It is defined precisely by

$$\mu = \left(\frac{\partial E}{\partial N}\right)_{V(r)} \qquad (9)$$

**Electronegativity**

$$\chi = -\mu \approx -\frac{E_{LUMO} + E_{HOMO}}{2} = \frac{EA + IP}{2} \qquad (10)$$

It characterizes a tendency to attract the electrons that make up the bond. In this process, an electron transferred from an atom or functional group of a molecule having a low ionization energy value IP can be completely transferred to a second functional group (atom) having a high electron affinity EA; as a result, an ionic bond will be formed. On the other hand, when these values are comparable, the common electron pair is shifted towards the functional group (atom) having a high value of χ, which is accompanied by the formation of a polarized bond. The electronegativity expression reproduces approximately the electronegativity of atoms and functional groups, the exact values are calculated from the formula:

$$\chi = -\left(\frac{\partial E}{\partial N}\right)_{V(r)} \qquad (11)$$

**Electrophilicity Index**

$$\omega \approx \frac{\chi^2}{2\eta} = \frac{(E_{LUMO} + E_{HOMO})^2}{4(E_{LUMO} - E_{HOMO})} \qquad (12)$$

It is an activity descriptor introduced by Robert Parr et al. (1999), which determines the energy change of an electrophilic reagent when it is saturated with electrons. An electrophilic reagent is a molecule or group of atoms in a molecule (acceptor) in which there is a deficit of electrons and the possibility of their attachment together with a carrier (donor). The higher value of ω implies that the compound can be considered a strong electrophile, hence a strong nucleophile is described by lower values of ω. The electrophilicity factor can be used as a measure of chemical and biological activity (Campodónico et al., 2005, 2008; Chattaraj et al., 2003; Chattaraj et al., 2006; Chattaraj and Giri, 2009; Enoch et al., 2008, 2009; Marinho et al., 2021; Parthasarathi, et al., 2004), in particular the toxicity of organic compounds. Analyzing the toxicity T(ω) of 37 polychlorinated dibenzofurans, expressed by the $pIC_{50}$=-log($IC_{50}$) inhibitory concentration parameter, Roy et al.,(2006) showed that it varied according to a linear equation T(ω)=a·ω+b, in which the parameters a=3.3944(0.3451), b=-5.5788(1.2804) were fitted with the least squares routine and an accuracy of $R^2$=0.7864; standard error of the estimate was SE=

0.7206. In more extended studies Shalini et al. (2017) proved that the toxicity $T(\omega)$ of aliphatic compounds such as: alcohols (amino alcohols, saturated, unsaturated, diols, halogenated alcohols, α-acetylene alcohols), esters, carboxylic acids, aldehydes, ketones and amines in total 252 tested substances, is described also by a linear relationship $T(\omega)=a\cdot\omega+b$ in which parameters a and b were fitted to the $pIC_{50}$ experimental data. Depending on the type of the compound analyzed, the following (a, b) parameters were determined: amino-alcohols (3.7676, -6.325), α-acetylated alcohols (2.9109, -2.9962), diols (28.852, -39.301), halogenated alcohols (3.0956, - 4.1069), saturated alcohols (71.829, -105.73), unsaturated alcohols (2.2531, -3.648), carboxylic acids (1.2124, -14752), halogenated carboxylic acids (2.7272, -2.6727), monoesters (8.0257, -9.5842), diesters ( 5.3498, -5.0322,), aldehydes (0.8757, -1.1151), ketones (18.489, -24.216), amines (-0.1421, -0.6487). The concept of the global ω electrophilicity has been extended (Chattaray et al., 2006) to include the local philicity, represented by the index $\omega_k^\alpha$ to be calculated by quantum-chemical computations as well as more and less extensive basis sets (Sanchez-Marquez et al., 2020). Here, k is the local (atomic) site in the molecule, whereas α=+,0,- for electrophilic, radical or nucleophilic reaction type.

In the present study, the calculations of the HOMO and LUMO energy levels and the values of global descriptors were carried out using Gaussian vs 16 software, the DFT B3LYP method and QZVP (Valence Quadruple-Zeta Polarization) basis set, which enable precise calculations for both light (H, Na, Ca) and heavy (Ag, Cu, Cd, Hg, K) atoms. The QZVP base can be used for molecules containing atoms from H to La and for Hf-Rn range in contrast to the popular basis, e.g. 6-311G, applied for compounds with H-Kr atoms or cc-pVDZ, cc-pVTZ, cc-pVQZ basis employed for molecules containing H-Ar and Ca-Kr atoms. From the quantum-chemical point of view, hydrogen cyanide and its salts are simple compounds endowed with nontrivial geometry. Therefore, in calculations, the non-linear structures of NaCN, KCN (T-shape) (Klein et al., 1981) and Ca(CN)$_2$ (twisted) (Kapp and Schleyer, 1996) have been taken into consideration. The remaining ones have linear form after the optimization procedure. The geometry of molecules optimized is presented in Tables A1 and A2 in Appendix. The test calculations performed for HCN, NaCN and KCN, by making use of DFT B3LYP and CCSD(T) methods as well as QZVP, cc-pVQZ, 6311++G(2df,2pd) basis sets revealed that the model applied in the calculations (DFT B3LYP QZVP) produces the lowest total energy for molecules considered (see Table 2).

Table 2. Total energy [Ha] of HCN, NaCN, KCN in the global minimum calculated using different quantum computational methods and basis sets.

| Method | HCN | NaCN | KCN |
|---|---|---|---|
| DFT B3LYP QZVP | -93.470347 | -255.210774 | -692.858177 |
| CCSD(T) QZVP | -93.301219 | -254.611922 | -691.922650 |
| CCSD(T) cc-pVQZ | -93.301011 | -255.208485 | |
| DFT B3LYP cc-pVQZ | -93.468628 | -255.208813 | |
| DFT B3LYP 6311++G(2df,2pd) | -93.459682 | -255.188599 | -692.835315 |

## 4. Results

From HOMO and LUMO energies calculated for the gas phase, the values of global descriptors characterizing the activity of hydrogen cyanide and its salts were determined – they are collected in Table 3. A detailed analysis of the results obtained reveals a correlation between $LD_{50}$ and EA, $\omega$-values for CuCN or AgCN, $Zn(CN)_2$, $Hg(CN)_2$, $Cd(CN)_2$, NaCN, KCN, HCN ordered according to the toxicity increased with diminishing EA, $\omega$-descriptors. In the case of the last three molecules, the mean values of $LD_{50}$ have been taken into interpretation. Such a correlation has not been observed for AgCN or CuCN), $Ca(CN)_2$ as well as for NaCN, KCN, represented by the most accurate values of $LD_{50}$ determined by Ballantyne and Salem (2008). To explain those inconsistencies and to obtain the most reliable toxicological characteristics of cyanides, calculations have been performed not only in gas phase but also in water medium using the C-PCM solvation model (Conductor–like Polarizable Continuum Model) (Cossi et al., 2003) implemented in software package Gaussian vs 16. Hydrolytic decay of cyanide salts is a key process responsible for their toxicity connected with sudden (NaCN, KCN) or gradual ($Ca(CN)_2$) HCN releasing. Hence, only the values of global descriptors presented in Table 4 should be treated as representative (realistic) for the compounds considered in contradistinction to the unrealistic results in Table 3. Inspection of Table 4 reveals correlation between EA and $LD_{50}$ values for CuCN or AgCN, $Hg(CN)_2$, $Cd(CN)_2$, KCN, NaCN, HCN as well as between $\omega$ and $LD_{50}$ adjusted as free cyanide CN moiety for CuCN or AgCN, $Hg(CN)_2$, $Cd(CN)_2$, HCN, KCN, NaCN in the order of increasing toxicity. Such a correlation is not found for $Zn(CN)_2$ and $Ca(CN)_2$, which behave as typical outliers with respect to the model under consideration. To express the above specified correlations quantitatively and to estimate the $LD_{50}$ for CuCN versus AgCN and $Zn(CN)_2$, $Ca(CN)_2$ consistent with the toxicity profile of the remaining cyanides, the analytical functions

$$T(EA) = T_0(EA_0)\exp\left\{[a(EA-EA_0)]^b\right\}, \qquad T(\omega) = T_0(\omega_0)\exp\left\{[a(\omega-\omega_0)]^b\right\} \qquad (13)$$

that represent the non-linear relationship between toxicity T and EA, ω-descriptors have been taken into consideration. From a mathematical point of view, Eqs.(13) represent the exponential part of the Avrami (1939, 1940, 1941) formula widely used in the description of phase changes, in particular the chemical reaction rates and kinetics of crystallization. Here, a [Ha$^{-1}$] and b are free adjustable parameters, fitted to the toxicity data from Table 2; T(EA) represents the theoretical values of LD$_{50}$(EA), EA$_0$=0.00221 [Ha] and T$_0$(EA$_0$) = LD$_{50}$(HCN) = 4.21 [mg/kg BW] are constrained to the values characterizing the most toxic HCN. T(ω) describes LD$_{50}^{CN}$(ω) expressed as a free CN$^-$ anion, ω$_0$=0.070643 [Ha] and T$_0$(ω$_0$) = LD$_{50}^{CN}$(NaCN) =3.04 [mg CN/kg BW] are constrained to the values characterizing NaCN toxicity. The parameters constrained ensure a proper behavior of the T(EA) and T(ω) functions in the limit of the highest toxicity (the smallest values of EA$_0$ and ω$_0$ parameters). The calculations have been performed by taking advantage of Sigma Plot vs 11 software - the results of calculations together with indicators of goodness of fit ($R^2$ – coefficient of determination, SD – standard deviation of parameters fitted, SE – standard error of estimation, F value of F-statistics) are presented in Table 5. The plots of the functions T(EA) and T(ω) for the best fits are presented in Fig.1a and Fig. 1b, respectively.

Table 5. The values of parameters fitted (a in [Ha$^{-1}$], b is dimensionless) to the toxicity functions T(EA) and T(ω) of cyanides X(CN)$_k$ for N data sets including X=H, Na, K, Cd, Hg (N=5), Ag and (or) Cu data. The values in parentheses are constrained - they represent EA$_0$ and ω$_0$ [Ha] calculated as well as T$_0$(EA$_0$) = LD$_{50}$ [mg/kg BW] for HCN and T$_0$(ω$_0$) = LD$_{50}^{CN}$ [mg CN/kg BW] for NaCN, obtained experimentally by Ballantyne and Salem (2008).

| N | Constrained | Fitted | SD | $R^2$ | SE | F |
|---|---|---|---|---|---|---|
| T(EA) | | | | | | |
| 5+Ag | T$_0$(EA$_0$) = [4.21] | a=83.5431 | 8.9439 | 0.9976 | 2.4982 | 1698 |
| | EA$_0$=[0.00221] | b= 0.7735 | 0.3329 | | | |
| 5+Cu | | a=68.9061 | 3.7291 | 1.0000 | 2.2056 | 269059 |
| | | b= 1.1339 | 0.0400 | | | |
| 5+Ag,Cu | | a=25.0705 | 0.6466 | 0.9996 | 10.7184 | 11332 |
| | | b= 3.3205 | 0.1646 | | | |
| T(ω) | | | | | | |
| 5+Ag | T$_0$(ω$_0$) = [3.04] | a=24.0772 | 0.6527 | 0.9976 | 0.4447 | 1672 |
| | ω$_0$=[0.070643] | b= 2.0796 | 0.1620 | | | |
| 5+Cu | | a=25.4655 | 0.8553 | 1.0000 | 0.4879 | 462158 |
| | | b= 2.5362 | 0.1378 | | | |
| 5+Ag,Cu | | a=20.4561 | 0.2225 | 0.9999 | 1.1432 | 85283 |
| | | b= 3.9298 | 0.1072 | | | |

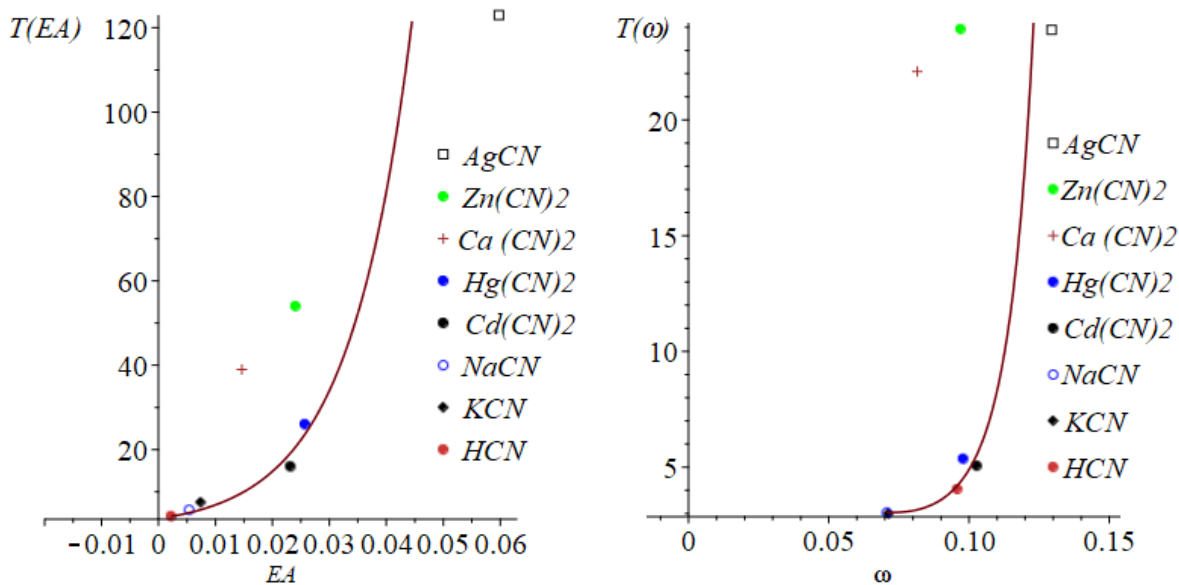

Fig. 1 a, b. Plots of the toxicity functions T(EA)=LD$_{50}$(EA), a=68.9061 [Ha$^{-1}$], b=1.1339 and T($\omega$)=LD$_{50}^{CN}$($\omega$), a=25.4655 [Ha$^{-1}$], b=2.5362 with parameters parameters from Table 5 fitted to HCN, NaCN, KCN, Cd(CN)$_2$, Hg(CN)$_2$,CuCN experimental data. CuCN data-point is not presented (out of scale), whereas AgCN, Ca(CN)$_2$, Zn(CN)$_2$ are not included in the fits (outliers).

The test calculations revealed that the presence of AgCN data-point in the fits deteriorates the reproduction of the total data set, hence it has been excluded from the fit together with Ca(CN)$_2$ and Zn(CN)$_2$ data being evident outliers. The toxicological characteristics of those three compounds are estimated from the functions T(EA)=LD$_{50}$(EA), T($\omega$)=LD$_{50}^{CN}$($\omega$) and presented in Table 6. Its inspection reveals that while the EA-parameter roughly determines the LD$_{50}$ for cyanides considered, the $\omega$-descriptor is related rather to the product of cyanide hydrolysis n-degree and its LD$_{50}$ according to the relationship:

$$LD_{50}^{CN} = \frac{M_{CN} \cdot n \cdot LD_{50}}{M_{X(CN)_k}} = T_0(\omega_0)\exp\left\{[a(\omega-\omega_0)]^b\right\} \qquad 0 \leq n \leq k = 1,2 \qquad (14)$$

Here, M$_{CN}$ and M$_{X(CN)k}$ denote the mole mass of the cyanide ion and the cyanide salt, which generates CN$^-$ by hydrolytic decay. Since n-parameter cannot be greater than the number of CN groups (n $\leq$ k = 1, 2) in the compound, the calculations performed justify the thesis that the LD$_{50}$ values determined experimentally for cyanides with n>k are incorrect (AgCN) or inaccurate (HCN, NaCN, KCN, Cd(CN)$_2$) and require revision. In particular, for AgCN the new value estimated is LD$_{50}$($\omega$)=253.72 [mg/kg BW], assuming 100% hydrolysis (n = 1). In the case of Zn(CN)$_2$ and Ca(CN)$_2$, the situation is different: the known experimental values 54 and 39 [mg/kg BW] are correct assuming a low degree of the first step of hydrolysis n = 0.3658 and 0.2870, respectively or they require a change to the values LD$_{50}$($\omega$)=9.88 and 5.60 [mg/kg BW], respectively assuming total first- and second-steps of hydrolysis (n=2). Comparison of the estimated $LD_{50}(\omega)$ with the NOAEL toxicity parameters for the cyanides considered (see Table 6), supports the second of the possibilities mentioned as the theoretical $LD_{50}(\omega)$ values are well correlated with the experimental NOAELs values in contrast to $LD_{50}(Exp)$ ones exhibiting only an approximate correlation. This conclusion can be proved by fitting $LD_{50}(Exp)$, $LD_{50}(EA)$, $LD_{50}(\omega)$ to NOAELs data available for HCN, NaCN, KCN, AgCN, $Zn(CN)_2$, $Ca(CN)_2$ (IRIS Integrated Risk Information System https://www.epa.gov/iris) by taking advantage of the function

$$LD_{50}(NOAEL) = LD_{50}^0 \exp\left\{\left[a(NOAEL - NOAEL_0)\right]^b\right\} \tag{15}$$

identical in the form as formulae (13). The results of calculations are presented in Table 7.

Table 7. The parameters of $LD_{50}(NOAEL)$ function (15) fitted (a in [mg/kg BW/day]$^{-1}$, b is dimensionless) to $LD_{50}(Exp)$, $LD_{50}(EA)$, $LD_{50}(\omega)$ and NOAELs data for N=6 cyanides $X(CN)_k$ for X=H, Na, K, Ag, Ca, Zn, reported in Table 6. The values of $LD^0_{50}$ and $NOAEL_0$ for HCN are constrained.

| N | Constrained | Fitted | SD | $R^2$ | SE | F |
|---|---|---|---|---|---|---|
| | | $LD_{50}(Exp)$ | | | | |
| 6 | $LD^0_{50}$= [4.21] | a=0.2827 | 0.2840 | 0.8129 | 22.2558 | 17 |
| | $NOAEL_0$=[10.8] | b=0.4778 | 0.1926 | | | |
| | | $LD_{50}(EA)$ | | | | |
| 6 | $LD^0_{50}$= [4.21] | a=0.0673 | 0.0174 | 0.9991 | 6.5974 | 4554 |
| | $NOAEL_0$=[10.8] | b=1.4126 | 0.3310 | | | |
| | | $LD_{50}(\omega)$ | | | | |
| 6 | $LD^0_{50}$=[4.31] | a=0.0517 | 0.0039 | 0.9998 | 1.4076 | 25684 |
| | $NOAEL_0$=[10.8] | b=1.6636 | 0.1505 | | | |

Inversion of the relation (15) to

$$NOAEL(LD_{50}) = NOAEL_0 + \frac{1}{a}\sqrt[b]{\ln\left(\frac{LD_{50}}{LD_{50}^0}\right)} \tag{16}$$

enables the theoretical estimation of NOAELs for cyanides for which they have not yet been determined. In particular, taking into account $LD_{50}(\omega)$ values presented in Table 6 for CuCN, $Hg(CN)_2$ and $Cd(CN)_2$ one gets NOAEL=65.69, 36.61, 34.37 [mg/kg BW/day], respectively.

## 5. Discussion

Analysis of the results presented in Table 5 confirms satisfactory for T(EA) and excellent for T($\omega$) correlations between experimental $LD_{50}$, $LD_{50}^{CN}$ values and EA, $\omega$-parameters treated as toxicity descriptors for cyanides X(CN)$_k$, X=H, Na, K, Cd, Hg, Cu. In particular, T(EA) confirms that the toxicity of hydrogen cyanide and its salts in water environment varies from the lowest to the highest in the order CuCN<Hg(CN)$_2$<Cd(CN)$_2$<KCN<NaCN<HCN, whereas T($\omega$) indicates the hierarchy CuCN<Hg(CN)$_2$<Cd(CN)$_2$<HCN<KCN<NaCN with respect to the molar concentration of the free CN$^-$ anion. While the EA-parameter roughly determines the LD$_{50}$ for cyanides considered, the $\omega$-descriptor is related to the product of cyanide hydrolysis n-degree and its LD$_{50}$. Hence, T($\omega$) toxicity function is interconnected with the n·LD$_{50}$ collective variable via relation (14), whereas T(EA) can approximate directly LD$_{50}$ values using equation (13). This conclusion indicates a key role of the degree of hydrolysis for the toxicity of cyanide salts connected with CN$^-$ production. It also supports a thesis that the toxicity of cyanide compounds depends on the tendency to release cyanide anion. It is well-known that cyanide radicals have a low affinity to alkali metals and a high affinity to other metals, especially the ferric cation Fe$^{3+}$. Hence, water solutions of cyanide salts are extremely toxic, in contrast to iron-containing cyanide complexes, which do not release CN$^-$ easy, consequently, they are practically non-toxic, even in the water medium. The LD$_{50}$(Exp) reported in Table 6 suggest a need to introduce a significant change in this parameter for AgCN and minor changes for HCN, NaCN, KCN and Cd(CN)$_2$. In the case of Zn(CN)$_2$ and Ca(CN)$_2$ the problem is open and requires experimental determination of the hydrolysis degrees of those compounds, which are currently unavailable (NRC, 2015). If we assume that the above cyanides undergo 100% hydrolysis in both the first- and second steps, the theoretically estimated LD$_{50}$($\omega$) values presented in Table 6 can be considered as representative. The calculations performed prove (see Tables 6 and 7) that they are consistent with the values of alternative toxicity parameters (NOAEL, AEGL-1) determined for the part of compounds investigated and allow cyanides to be ranked with respect to the lowest up to the highest toxicity in the following manner:

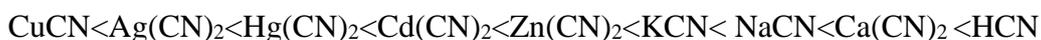
CuCN<Ag(CN)$_2$<Hg(CN)$_2$<Cd(CN)$_2$<Zn(CN)$_2$<KCN< NaCN<Ca(CN)$_2$ <HCN

## 6. Conclusions

Although the LD$_{50}$ toxicity parameter is nowadays considered archaic and replaced by other modern descriptors such as AEGL, BMD, BMDL, NTEL, NOEL, NOAEL, MTD, the LD$_{50}$ values determined so far are still present in various databases characterizing the toxicity of chemical compounds (ATSDR, 1989; MAK, 2003; NRC, 2015; RTECS, 2015). The results obtained in this study indicate that the LD$_{50}$ values for cyanides can be satisfactory reproduced by quantum-chemical calculations using the $\omega$-toxicity descriptor. In this way, one can select the most reliable LD$_{50}$ values from a wide range of available experimental data and verify their correctness. The EA-parameter only roughly determines the $LD_{50}$ for the cyanides considered, whereas the ω-descriptor is related to the product of n-degree hydrolysis of cyanide salts and their $LD_{50}$ via a non-linear relationship (14). A similar relation links the theoretically estimated values of $LD_{50}(ω)$ with experimental NOAELs and can be used to compute the unknown NOAEL value when $LD_{50}$ is known. The ω-descriptor can be used to quantify the toxicity of not only organic compounds, as has already been proven, but also inorganic ones, in particular cyanides, which are classified as highly toxic substances. Their toxicity decreases in a non-linear manner with increasing EA, ω-values, in contrast to the organic compounds so far investigated, whose toxicity decreased linearly with ω-parameter. QSAR model applied in this study is based on the HOMO, LUMO energy levels and global EA, ω-descriptors, confirming their important role in characterizing the acute toxicity of substances in general and cyanides in particular.

**Literature**

Al-Fahemi. J.H., 2012. The use of quantum-chemical descriptors for predicting the photoinduced toxicity of PAHs. J Mol Model 18, 4121-9. DOI: 10.1007/s00894-012-1417-0.

Arwa, B.R., Vladimir, B.B., 2016. In silico toxicology: computational methods for the prediction of chemical toxicity. Wiley Interdiscip Rev Comput Mol Sci. 6, 147–172. DOI: 10.1002/wcms.1240

ATSDR, 1989. Agency for Toxic Substances and Disease Registry. Toxicological Profile for Cyanide. ATSDR/TP-88/12; PB90-162058. Prepared by Syracuse Research Corporation for ATSDR, U.S. Public Health Service, under Contract No. 68-C8-0004.
https://www.atsdr.cdc.gov

Avrami, M., 1939. Kinetics of Phase Change. I. General Theory. Journal of Chemical Physics. 7, 1103–1112. DOI:10.1063/1.1750380.

Avrami, M., 1940. Kinetics of Phase Change. II. Transformation-Time Relations for Random Distribution of Nuclei. Journal of Chemical Physics. 8, 212–224. DOI:10.1063/1.1750631.

Avrami, M., 1941. Kinetics of Phase Change. III. Granulation, Phase Change, and Microstructure. Journal of Chemical Physics. 9, 177–184.. DOI:10.1063/1.1750872

Ballantyne, B., 1988. Toxicology and hazard evaluation of cyanide fumigation powders. Clin. Toxicol. 26, 325–335. DOI: 10.1080/15563658809167097

Ballantyne, B., Salem, H.,2008. Cyanides: Toxicology, Clinical Presentation, and Medical Management. In Chemical Warfare Agents Chemistry, Pharmacology, Toxicology and Therapeutics. Edited by Romano J.A., Lukey, B.J., Salem, H. Second Edition. Taylor & Francis Group. ISBN: 9781420046618

Campodónico, P.R., Contreras, R., 2008. Structure–reactivity relationships for electrophilic sugars in interaction with nucleophilic biological targets. Bioorg. Med. Chem. 16, 3184-3190. DOI: 10.1016/j.bmc.2007.12.018.

Campodónico, P.R., Fuentealba, P., Castro, E.A., Santos, J.G., Contreras, R., 2005. Relationships between the electrophilicity index and experimental rate coefficients for the aminolysis of thiolcarbonates and dithiocarbonates. J. Org. Chem. 70, 1754-1760. DOI: 10.1021/jo048127k.

Chattaraj, P. K., Maiti, B., Sarkar, U., 2003. Philicity: unified treatment of chemical reactivity and selectivity. J. Phys. Chem. 107, 4973−4975. https://doi.org/10.1021/jp034707u

Chattaraj, P.K., Sarkar, U., Roy, D. R. 2006. Electrophilicity index. Chem. Rev. 106, 2065−2091. https://doi.org/10.1021/cr040109f

Chattaraj, P. K., Giri, S., 2009. Electrophilicity index within a conceptual DFT framework. Annu. Rep. Prog. Chem., Sect. C: Phys. Chem.105, 13−39. https://doi.org/10.1039/B802832J

Cherkasov, A., Muratov, E. N., Fourches, D., Varnek, A., Baskin, I. I., Cronin, M., Dearden, J. C., Gramatica, P., Martin, Y. C., Todeschini, R., Consonni, V., Kuz, V. E., Cramer, R. D., Benigni, R., Yang, C., Rathman, J. F., Terfloth, L., Gasteiger, J., Richard, A. M.,Tropsha, A., 2014. QSAR modeling: Where have you been? Where are you going to? Journal of Medicinal Chemistry, 57, 4977–5010.v https://doi.org/10.1021/jm4004285

Christopher, P., Holstege, Ch.P., 2010. Criminal Poisoning: Clinical and Forensic Perspectives. Jones & Bartlett Publishers. ISBN-13: 978-0763744632

Cossi, M., Rega, N., Scalmani, G., Barone, V., 2003. Energies, Structures, and Electronic Properties of Molecules in Solution with the C-PCM Solvation Model. Journal of Computational Chemistry 24, 669–681. DOI: 10.1002/jcc.10189

Cronin, L.T.D., 2002. The current status and future applicability of quantitative structure-activity relationships (QSARs) in predicting toxicity. Altern Lab Anim. 2, 81-4. DOI: 10.1177/026119290203002S12.

Dearden, J.C., 2003. In silico prediction of drug toxicity. J Comput Aided Mol Des.17, 119-27. DOI: 10.1023/a:1025361621494.

Deeb, O., Goodarzi, M., 2012. In silico quantitative structure toxicity relationship of chemical compounds: some case studies. Curr Drug Saf 7, 289–297. DOI: 10.2174/157488612804096533

Egekeze, J.O., Oehme, F.W., 1980. Cyanides and their toxicity: A literature review, Veterinary Quarterly, 2, 104-114, DOI: 10.1080/01652176.1980.9693766

Enoch, S.J., Cronin, M.T.D., Schultz, T.W., Madden, J.C., 2008. Quantitative and mechanistic read across for predicting the skin sensitization potential of alkene acting via Michael addition. Chem. Res. Toxicol. 21, 513-520. DOI: 10.1021/tx700322g

Enoch, S.J., Roberts, D.W., Cronin, M.T.D., 2009. Electrophilic reaction chemistry of low molecular weight respiratory sensitizers. Chem. Res. Toxicol. 22, 1447-1453. DOI: 10.1021/tx9001463

Faust, R.A., 1994. The Risk Assessment Information System. Formal Toxicity Summary for cyanide. Chemical Hazard Evaluation and Communication Group, Biomedical and Environmental Information Analysis Section, Health and Safety Research Division, Oak Ridge, Tennessee. https://rais.ornl.gov

Forst, J.A., 1928. Zur Entgiftung der Blausäure (Detoxification of hydrogen cyanide) (German). Naunyn-Schmiedebergs Arch Pharmakol 128, 1–66

Gaines, T.B., 1969. Acute toxicity of pesticides. Toxicol Appl Pharmacol 14, 515-534. DOI: 10.1016/0041-008x(69)90013

Guha, R., Willighagen, E.L., 2012. A survey of quantitative descriptions of molecular structure. Current Topics in Medicinal Chemistry, 12, 1946–1956. DOI:10.1016/j.biotechadv.2011.08.021.Secreted

Hardy, A., et al., 2017. Update: use of the benchmark dose approach in risk assessment. EFSA Journal 15, 4658 (1-40). DOI: https://doi.org/10.2903/j.efsa.2017.4658

Howard, J.W., Hanzal, R.F., 1955. Chronic toxicity for rats by food treated with hydrogen cyanide. Agric. Food Chem. 3, 325-329. https://doi.org/10.1021/jf60050a004

Kapp, J., Schleyer, P.R., 1996. M(CN)2 Species (M = Be, Mg, Ca, Sr, Ba): Cyanides, Nitriles, or Neither? Inorg. Chem. 35, 2247–2252. https://doi.org/10.1021/ic9511837

Karelson, M., 2000. Molecular Descriptors in QSAR/QSPR. Wiley-InterScience, New Jersey. ISBN: 978-0-471-35168-9

Kleandrova, V.V., Luan, F., Speck-Planche, A., Cordeiro, M.N., 2015. In silico assessment of the acute toxicity of chemicals: recent advances and new model for multitasking prediction of toxic effect. Mini Rev Med Chem 15, 677-86. DOI: 10.2174/1389557515666150219143604.

Klein, M.L., Goddard, J.D., Bounds, D.G., 1981. An ab initio molecular orbital study of NaCN and KCN. J. Chem. Phys.75, 3909-3915. https://doi.org/10.1063/1.442547

Lorke, D., 1983. A new approach to practical acute toxicity testing. Arch Toxicol 54, 275-287. DOI: 10.1007/BF01234480.

MAK, 2003. Collection for Occupational Health and Safety: Annual Thresholds and Classifications for the Workplace. Hydrogen cyanide, potassium cyanide and sodium cyanide. Wiley Online Library. https://doi.org/10.1002/3527600418.mb7490vere0019

Marinho, M.M. et al., 2021. Quantum computational investigations and molecular docking studies on amentoflavone. Heliyon 7, e06079. DOI: 10.1016/j.heliyon.2021.e06079

NRC, 2015. National Research Council (US) Committee on Acute Exposure Guideline Levels. Acute Exposure Guideline Levels for Selected Airborne Chemicals: Vol. 19. Chapter:

1 Cyanide Salts Acute Exposure Guideline Levels. Washington (DC), National Academies Press (US). ISBN-13: 978-0-309-14515-2

Parr, R.G., Szentpály, L.V., Liu, S., 1999. Electrophilicity Index. J. Am. Chem. Soc. 121, 1922-1924. https://doi.org/10.1021/ja983494x

Parr, R.G., Yang, W., 1989. Density Functional Theory of Atoms and Molecules. Oxford University Press, Oxford. ISBN-13: 978-0-195-09276-9

Parthasarathi, R., Subramanian, V., Roy, D.R., Chattaraj, P.K., 2004. Electrophilicity index as a possible descriptor of biological activity. Bioorganic & Medicinal Chemistry 12, 5533-5543. DOI: 10.1016/j.bmc.2004.08.013

Pearson, R.G. 1997. Chemical Hardness - Applications from Molecules to Solids. Wiley-VCH, Weinheim. ISBN: 978-3-527-60617-7

Philbrick, D.J., Hopkins, J.B., Hill, D.C., Alexander, J.C., Thomson, R.G., 1979. Effects of prolonged cyanide and thiocyanate feeding in rats. J. Toxicol. Environ. Health 579-592. DOI: 10.1080/15287397909529770.

Roy, D.R, Sarkar, U., Chattaraj, P.K., Mitra, A., Padmanabhan, J., Parthasarathi, R., Subramanian, V., Van Damme, S., Bultinck, P., 2006. Analyzing toxicity through electrophilicity. Molecular Diversity 10, 119–131. DOI: 10.1007/s11030-005-9009-x

RTECS, 2015. Registry of Toxic Effects of Chemical Substances. National Institute for Occupational Safety and Health, Ohio, Cincinnati. https://www.cdc.gov

Sanchez-Marquez, J., García,V., Zorrilla, D., Fernandez, M., 2020. On Electronegativity, Hardness, and Reactivity Descriptors: A New Property-Oriented Basis Set. J. Phys. Chem. 124, 4700−4711. https://doi.org/10.1021/acs.jpca.0c01342

Shalini A., Tandon, H., Chakraborty T., 2017. Molecular Electrophilicity Index – A Promising Descriptor for Predicting Toxicological Property. Journal of Bioequivalence & Bioavailability106, 2065–209. DOI:10.4172/jbb.1000356

Smyth, H.F., Carpenter, C.P., Weil, C.S., Pozzani, U.C., Nycum, J.C., 1969. Range-finding toxicity data: list VII. American Industrial Hygiene Association Journal, 30(5):470-476. DOI: 10.1080/00028896909343157

Sterner, RT., 1979. Effects of sodium cyanide and diphacinone in coyotes (*canis latrans*): applications as predacides in livestock toxic collars. Bulletin Environ Contam Toxicol 23, 211-217. DOI: 10.1007/BF01769944

Talmaciu, M.M., Bodoki, E., Oprean, R., 2016. Global chemical reactivity parameters for several chiral beta-blockers from the Density Functional Theory viewpoint. Clujul Med. 89, 513–518. DOI: 10.15386/cjmed-610

Todeschini, R., Consonni, V., 2009. Molecular Descriptors for Chemoinformatics: Volume I: Alphabetical Listing/Volume II: Appendices, References, Volume 41. Wiley-VCH Verlag GmbH & Co. DOI:10.1002/9783527628766

Valerio, L.G., 2009. In silico toxicology for the pharmaceutical sciences. Toxicol Appl Pharmacol. 241, 356-70. DOI: 10.1016/j.taap.2009.08.022.

Vikas, R., 2015. Exploring the role of quantum chemical descriptors in modeling acute toxicity of diverse chemicals to Daphnia magna. Journal of Molecular Graphics and Modelling 61, 89-101. https://doi.org/10.1016/j.jmgm.2015.06.009

Table 3. The values of global descriptors of cyanides in gas phase, calculated from the energies $E_{LUMO}=-EA$ and $E_{HOMO}=-IP$ determined at DFT B3LYP QZVP theory level. The compounds are ordered with respect to decreasing experimental $LD_{50}$ and increasing toxicity.

| Compound | EA [Ha] | IP [Ha] | ΔE [eV] | η [Ha] | S [Ha$^{-1}$] | μ=-χ [Ha] | ω [Ha] | $LD_{50}$ [mg/kg BW] |
|---|---|---|---|---|---|---|---|---|
| CuCN | 0.12874 | 0.29525 | 4.5310 | 0.08326 | 6.0056 | -0.2119 | 0.26991 | 1265.00 |
| AgCN | 0.13360 | 0.30340 | 4.6205 | 0.08490 | 5.8889 | -0.2185 | 0.28117 | 123.00 |
| Zn(CN)$_2$ | 0.09852 | 0.36703 | 7.3065 | 0.13426 | 3.7243 | -0.2327 | 0.20180 | 54.00 |
| Ca(CN)$_2$ | 0.09175 | 0.30448 | 5.7887 | 0.10637 | 4.7008 | -0.1981 | 0.18450 | 39.00 |
| Hg(CN)$_2$ | 0.09702 | 0.36409 | 7.2674 | 0.13349 | 3.7456 | -0.2305 | 0.19903 | 26.00 |
| Cd(CN)$_2$ | 0.09413 | 0.36067 | 7.2529 | 0.13327 | 3.7518 | -0.2274 | 0.19401 | 16.00 |
| NaCN | 0.07226 | 0.24461 | 4.6899 | 0.08618 | 5.8021 | -0.1584 | 0.14564 | 8.05[a] |
| KCN | 0.05953 | 0.22642 | 4.54131 | 0.08345 | 5.9920 | -0.1429 | 0.12249 | 7.64[a] |
| HCN | 0.00597 | 0.37452 | 10.0288 | 0.18428 | 2.7133 | -0.1902 | 0.09820 | 4.11[a] |

[a] Mean value reported in Table 1.

Table 4. The values of global descriptors of cyanides in the water medium, determined from the energies $E_{LUMO}=-EA$ and $E_{HOMO}=-IP$ calculated at the DFT B3LYP QZVP theory level using C-PCM solvation model. The $LD_{50}$ values are predicted from the functions $T(\omega)=LD_{50}^{CN}(\omega)$, $T(EA)=LD_{50}(EA)$ and the parameters obtained from the best fits, specified in Table 5.

| Compound | EA [Ha] | IP [Ha] | ΔE [eV] | η [Ha] | S [Ha$^{-1}$] | χ=- μ [Ha] | ω [Ha] | $LD_{50}^{CN}$(Exp)[a] [mg CN/kg BW] | $LD_{50}^{CN}$(ω) [mg CN/kg BW] | $LD_{50}$(Exp)[a] [mg/kg BW] | $LD_{50}$(EA) [mg/kg BW] |
|---|---|---|---|---|---|---|---|---|---|---|---|
| Fitted | | | | | | | | | | | |
| CuCN | 0.06962 | 0.26679 | 5.3653 | 0.0986 | 5.0718 | 0.1682 | 0.143495 | 367.38 | 367.19 | 1265.00 | 1267.51 |
| Hg(CN)$_2$ | 0.02570 | 0.30506 | 7.6018 | 0.1397 | 3.5796 | 0.1654 | 0.097904 | 5.35 | 4.52 | 26.00 | 23.66 |
| Cd(CN)$_2$ | 0.02315 | 0.33457 | 8.4742 | 0.1557 | 3.2111 | 0.1789 | 0.102727 | 5.06 | 5.53 | 16.00 | 19.16 |
| KCN | 0.00742 | 0.26200 | 6.9275 | 0.1273 | 3.9280 | 0.1347 | 0.071281 | 2.99 | 3.04 | 7.49 | 5.76 |
| NaCN | 0.00536 | 0.26605 | 7.0937 | 0.1303 | 3.8360 | 0.1357 | 0.070643 | 3.04 | 3.04 | 5.72 | 5.02 |
| HCN | 0.00221 | 0.37656 | 10.1866 | 0.1872 | 2.6713 | 0.1894 | 0.095811 | 4.05 | 4.20 | 4.21 | 4.21 |
| Predicted | | | | | | | | | | | |
| AgCN | 0.05976 | 0.27091 | 5.7457 | 0.1056 | 4.7360 | 0.1653 | 0.129461 | 23.89 | 49.31 | 123.00 | 496.55 |
| Zn(CN)$_2$ | 0.02405 | 0.30773 | 7.7193 | 0.1418 | 3.5251 | 0.1659 | 0.097009 | 23.92 | 4.38 | 54.00 | 20.64 |
| Ca(CN)$_2$ | 0.01463 | 0.27913 | 7.1974 | 0.1322 | 3.7807 | 0.1469 | 0.081564 | 22.09 | 3.16 | 39.00 | 8.75 |

[a] Experimental data reported in Table 1 including HCN, NaCN and KCN data evaluated by Ballantyne and Salem (2008).

Table 6. Theoretical and experimental toxicity characteristics of hydrogen cyanide and its salts $X(CN)_k$, k=1, 2, compared with experimental values of $LD_{50}(Exp)$, NOAEL and AEGL-1. The hydrolysis degree n-parameter reveals the correct (n≤k) and incorrect (n>k) values of experimental $LD_{50}(Exp)$ and theoretical $LD_{50}(EA)$ determinations.

| Compound | $LD_{50}(Exp)$ [mg/kg BW] | n | $LD_{50}(EA)$ [mg/kg BW] | n | $LD_{50}(\omega)$ [mg/kg BW] | n | n·$LD_{50}(\omega)$ [mg/kg BW] | NOAEL[a] [mg/kg BW/day] | AEGL-1[b] [mg/m$^3$/min] |
|---|---|---|---|---|---|---|---|---|---|
| | | | | | Fitted | | | | |
| CuCN | 1265.00 | 0.9992 | 1267.51 | 0.9973 | 1264.03 | 1 | 1264.03 | | |
| Hg(CN)$_2$ | 26.00 | 1.6879 | 23.66 | 1.8546 | 21.94 | 2 | 43.88 | | |
| Cd(CN)$_2$ | 16.00 | 2.1861 | 19.16 | 1.8253 | 17.49 | 2 | 34.98 | | |
| KCN | 7.49 | 1.0158 | 5.76 | 1.3218 | 7.61 | 1 | 7.61 | 27.0 | 6.6 |
| NaCN | 5.72 | 1.0011 | 5.02 | 1.1397 | 5.73 | 1 | 5.73 | 20.4 | 5.0 |
| HCN | 4.21 | 1.0367 | 4.21 | 1.0367 | 4.36 | 1 | 4.36 | 10.8 | |
| | | | | | Predicted | | | | |
| AgCN | 123.00 | 2.0628 | 496.55 | 0.5110 | 253.72 | 1 | 253.72 | 55.7 | |
| Zn(CN)$_2$ | 54.00 | 0.3658 | 20.64 | 0.9573 | 9.88 | 2 | 19.75 | 24.3 | |
| Ca(CN)$_2$ | 39.00 | 0.2870 | 8.75 | 1.2796 | 5.60 | 2 | 11.19 | 19.1 | 4.7 |

[a]IRIS Integrated Risk Information System https://www.epa.gov/iris

[b]NRC (2015)

# Appendix

Table A1. Optimized structures of hydrogen cyanide and its salts in the gas phase and the global energetic minimum E [Ha] calculated at the DFT B3LYP QZVP theory level.

| | | |
|---|---|---|
| 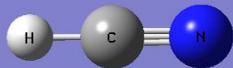 | 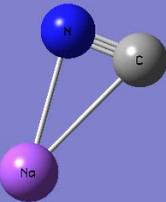 | 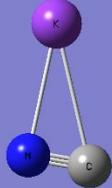 |
| HCN E=-93.470347 | NaCN E=-255.210775 | KCN E=-692.858177 |
| 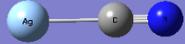 | 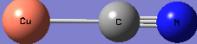 | 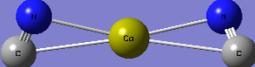 |
| AgCN E=-239.918672 | CuCN E=-1733.492878 | Ca(CN)$_2$ E=-863.444021 |
| 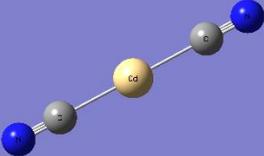 | 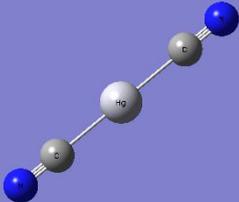 | 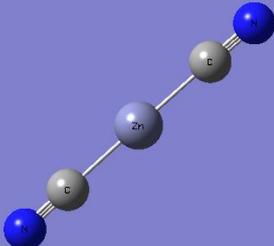 |
| Cd(CN)$_2$ E=-353.592754 | Hg(CN)$_2$ E=-399.299600 | Zn(CN)$_2$ E=-1965.268337 |

Table A2. Optimized structures of hydrogen cyanide and its salts in the water medium and the global energetic minimum E [Ha] calculated at the DFT B3LYP QZVP theory level and C-PCM solvation model.

| 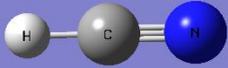 | 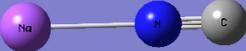 | 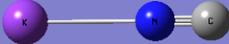 |
|---|---|---|
| HCN E=-93.477593 | NaCN E=-255.279127 | KCN E=-692.915850 |
| 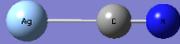 | 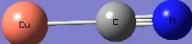 | 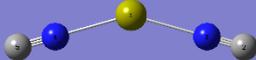 |
| AgCN E=-239.959900 | CuCN E=-1733.523161 | Ca(CN)$_2$ E=-863.547428 |
| 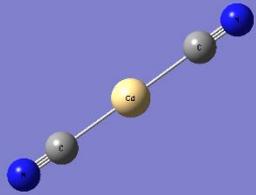 | 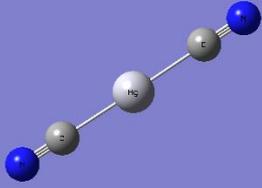 | 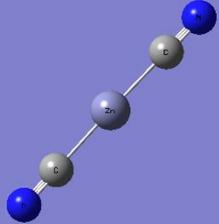 |
| Cd(CN)$_2$ E=-353.667549 | Hg(CN)$_2$ E=-399.369670 | Zn(CN)$_2$ E=-1965.329403 |